\documentclass{ifacconf}

\usepackage{natbib}        
\usepackage[pdftex]{graphicx}
\usepackage{amsmath,amssymb}
\usepackage{times}
\usepackage{bm}
\usepackage{here}
\usepackage{ascmac}
\usepackage{array}
\usepackage{longtable}
\usepackage{float}
\usepackage{wrapfig}
\usepackage{enumerate}
\usepackage{url}
\usepackage{multirow}
\usepackage{comment}
\usepackage{listings}
\usepackage{empheq}

\newcommand{\rfig}[1]{Fig.~\ref{#1}} 
\newcommand{\rfigs}[1]{Figs.~\ref{#1}}
\renewcommand{\figurename}{Fig. }
\renewcommand{\tablename}{Table } 

\newcommand{\Acal}{{\cal A}} 
\newcommand{\Fcal}{{\cal F}} 
\newcommand{\Kcal}{{\cal K}} 
\newcommand{\Scal}{{\cal S}}

\newtheorem{lemm}{Lemma}
\newtheorem{defi}{Definition}
\newtheorem{problem}{Problem}

\begin{document}
\begin{frontmatter}

\title{Learning Koopman Operator under Dissipativity Constraints}

\author[First]{Keita~Hara}, 
\author[First]{Masaki~Inoue},
\author[Second]{Noboru~Sebe} 

\address[First]{Department of Applied Physics and Physico-Informatics, Keio University,\\ 3-14-1 Hiyoshi, Kohoku-ku, Yokohama, Kanagawa, Japan.}
\address[Second]{Department of Artificial Intelligence, Kyushu Institute of Technology,\\ 680-4 Kawazu, Iizuka, Fukuoka, Japan.}

\begin{abstract}                
This paper addresses a learning problem for nonlinear dynamical systems with incorporating any specified dissipativity property.   The nonlinear systems are described by the Koopman operator, which is a linear operator defined on the infinite-dimensional lifted state space.  The problem of learning the Koopman operator under specified quadratic dissipativity constraints is formulated and addressed.  The learning problem is in a class of the non-convex optimization problem due to nonlinear constraints and is numerically intractable.  By applying the change of variable technique and the convex overbounding approximation, 
the problem is reduced to sequential convex optimization and is solved in a numerically efficient manner.  Finally, a numerical simulation is given, where high modeling accuracy achieved by the proposed approach including the specified dissipativity is demonstrated. 
\end{abstract}

\begin{keyword}
Learning, Dissipativity, Koopman Operator, Linear Matrix Inequality
\end{keyword}

\end{frontmatter}

\section{Introduction}
Artificial neural networks have re-developed recently and have brought high modeling accuracy in data-driven modeling of complex nonlinear systems.
In the development, multi-layered hierarchical models play a central role, 
and their efficient learning methods receive considerable attention  in various areas
(see e.g., the works by \cite{sitompul2013adaptive},  \cite{de2016randomized}, \cite{jin2016modeling}).
For example, \cite{de2016randomized} presents a identification method for nonlinear dynamical systems by using deep learning techniques.
\cite{jin2016modeling} proposes a deep reconstruction model (DRM) to analyze the characteristics of nonlinear systems.




The drawback of such learning methods, in particular for researchers or engineers, 
is on the gap between the constructed model and some {\it a priori} information on a physical system.  
Although the model may fit a given data-set and emulate the system behavior accurately, 
it cannot necessarily possess some practically essential properties of the system.
If we know {\it a priori} information on the properties, it is natural to try to incorporate them into the model.
For {\it linear} dynamical systems, a variety of learning methods that incorporate  {\it a priori} information have been studied well.
For example, the subspace identification method is combined with the {\it a priori} information including
stability  by  \cite{lacy2003subspace},
eigenvalue location by  \cite{okada1996subspace,miller2013subspace},
steady-state property by \cite{alenany2011improved,yoshimura2019system},
moments by \cite{inoue2019subspace}, and
positive-realness studied by \cite{goethals2003identification,hoagg2004first}, 
more general frequency-domain property by \cite{abe2016subspace}.
However, to the best of the authors' knowledge, for {\it nonlinear} dynamical systems, learning methods with {\it a priori} information have not been studied well.

This paper addresses a learning problem for nonlinear dynamical systems with incorporating the {\it a priori} information on {\it dissipativity}, 
which is proposed by \cite{Willems1972} and developed e.g., by \cite{Hill_76,Hill_77}.
To this end, the {\it nonlinear} system is described with the {\it Koopman operator}, 
which is a {\it linear} operator defined on infinite-dimensional lifted state space and has been applied to analysis of nonlinear dynamical systems.
See e.g., the works by \cite{koopman1931hamiltonian}, 
\cite{kevrekidis2015kernel}, \cite{williams2015data}, 
\cite{korda2018linear}.
Then, the learning problem is reduced to the data-driven finite-dimensional approximation of the Koopman operator onto the dissipativity constraint.

The approximation is formulated as 
the minimization of a convex cost function, which measures the consistency of the model and data, 
subject to a nonlinear matrix inequality, which represents the dissipation inequality.
The formulated problem, which is called Problem \ref{prob:1}, is in a class of the non-convex optimization and is numerically intractable.
Therefore, we aim at approximating Problem \ref{prob:1} to derive a numerically efficient algorithm that is composed of the solutions to the following two problems.
1) By applying the change of variable technique to the nonlinear matrix inequality, we derive a linear matrix inequality (LMI) constraint.
In addition, the approximation of the cost function reduces Problem \ref{prob:1} to a convex optimization, which is called Problem 2 and provides the feasible solution to Problem \ref{prob:1}. 
2) The convex overbounding approximation method proposed by \cite{sebe2018sequential} is applied to the nonlinear matrix inequality of Problem \ref{prob:1} 
to derive its inner approximation.  Then, the derived convex optimization problem, which is called Problem \ref{prob:3}, is sequentially solved by starting from the initial guess obtained in Problem \ref{prob:2}. 
It is guaranteed that the overall algorithm generates a less conservative solution than the solution obtained in Problem \ref{prob:2}.


The remaining parts of this paper is organized as follows.  In Section \ref{sec2}, we review theory of the Koopman operator and dissipativity.
In Section \ref{sec3}, the problem of learning the Koopman operator with the dissipativity-constraint is formulated, and the learning algorithm is proposed.
In Section \ref{sec4}, a numerical simulation is performed, in which the learning problem of a dissipative nonlinear system is addressed and the effectiveness of the algorithm is presented.  
Section \ref{sec5} concludes the works in this paper. 
\section{Preliminaries}
\label{sec2}

\subsection{Koopman Operator Theory}

In this section, we review {\it Koopman operator theory} to show its application to control systems based on the work by \cite{korda2018linear}.

\subsubsection{Koopman Operator}

We consider a discrete-time nonlinear system described by
\begin{align}
\left\{
	\begin{array}{ll}
		x(k+1) &= f\left(x(k),u(k)\right),\\
		y(k)&= g\left(x(k)\right),
	\label{eq.a0}
	\end{array}
\right.
\end{align}
where $k$ is the discrete time, $u\in \mathbb{R}^m$ is the input, $x\in \mathbb{R}^n$ is the state, $y\in \mathbb{R}^l$ is the output, 
and $f(x,u):\mathbb{R}^{n+m} \to \mathbb{R}^{n}$ and $g(x):\mathbb{R}^{n} \to \mathbb{R}^{l}$ are the nonlinear functions.
Let $z$ denote the extended state
\begin{align}
z:=
\left[\begin{array}{c}
x\\
u
\end{array}
\right]
\in \mathbb{R}^{n+m}.
\nonumber
\end{align}
Further let $\Fcal$ be the nonlinear operator defined by 
\begin{align}
\Fcal (z)&:=
	\left[\begin{array}{c}
	f(x,u)\\
	\Scal (u)
	\end{array}\right],
\nonumber
\end{align}
where $\Scal$ is the time-shift operator defined as
\begin{align}
\Scal (u(k))&:=u(k+1).
\nonumber
\end{align}
Then, the time evolution of $z$ is described by
\begin{align}
z(k+1)=\Fcal (z(k)).
\label{eq.a2}
\end{align}

\begin{figure}[t]
\centering
\includegraphics[scale=0.3]{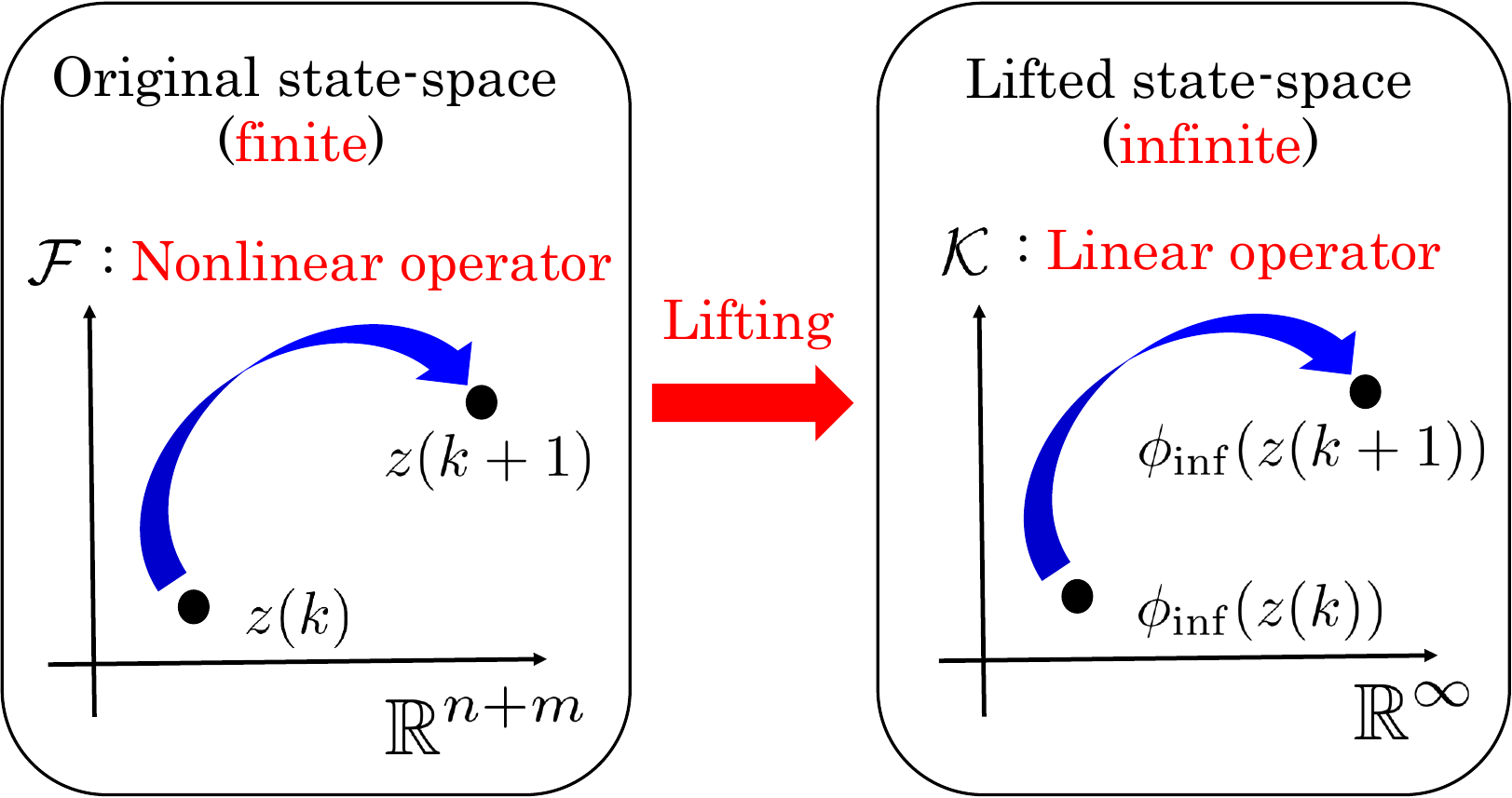}
\caption{Nonlinear operator $\mathcal{F}$ on state space and Koopman operator $\mathcal{K}$ on lifted state space}
\label{fig_koop}
\end{figure}

Now, we let $\phi_{\mathrm{inf}}(z)$ denote the infinite-dimensional lifting function described by
\begin{align}
\phi_{\mathrm{inf}}(z)=\left[\begin{array}{c}
\phi_{1}(z)\\
\phi_{2}(z)\\
\vdots
\end{array}
\right].
\end{align}
Here, we introduce the Koopman operator $\mathcal{K}$ as
\begin{align}
\mathcal{K}(\phi_{\mathrm{inf}}(z)):=\phi_{\mathrm{inf}}(\Fcal (z)).
\nonumber
\end{align}
Then, the time evolution of $\phi_{\mathrm{inf}}(z)$ is described by 
\begin{align}
\phi_{\mathrm{inf}}(z(k+1))=\mathcal{K}(\phi_{\mathrm{inf}}(z(k))).
\label{eq.a3.1}
\end{align}
Note that the Koopman operator is a linear operator defined on the infinite-dimensional state space, 
while expressing nonlinear dynamical systems (see e.g., the work by \cite{korda2018linear}).
\rfig{fig_koop} gives a sketch of the nonlinear operator $\Fcal$ on the state space and the Koopman operator $\Kcal$ on the lifted state space.


\subsubsection{{Approximation of Koopman Operator}}
Since the Koopman operator is the infinite-dimensional operator, it is difficult to be handled in numerical calculations.
In this subsection, we give the finite-dimensional approximation of the Koopman operator.
To this end, we define 
the $N_{\phi}$-dimensional lifting function $\phi(z):\mathbb{R}^{n+m} \to \mathbb{R}^{N_{\phi}}$ as
\begin{align}
\phi(z)=\left[\begin{array}{c}
\phi_{1}(z)\\
\vdots\\
\phi_{N_{\phi}}(z)
\end{array}
\right]
\in \mathbb{R}^{N_{\phi}}.
\nonumber
\end{align}
Furthermore, we let $\mathcal{A}\in \mathbb{R}^{N_{\phi}\times N_{\phi}}$ be a finite-dimensional matrix that approximates the Koopman operator $\mathcal{K}$, i.e., the error
\begin{align}
\|\mathcal{A}\phi(z) - \phi(\mathcal{F}(z))\|
\label{eq.a3.3}
\end{align}
is sufficiently small.
With this $\mathcal{A}$, we have the expression 
\begin{align}
\phi(z(k+1)) \approx \mathcal{A}\phi(z(k)),
\label{eq.a3}
\end{align}
which approximately describes the behavior of $\phi_{\mathrm{inf}}(z)$, defined by (\ref{eq.a3.1}).

In this paper we propose the method of the data-driven approximation of $\Kcal$, i.e., the learning method of $\Acal$ by using some data.
In the method, we aim at constructing the model (\ref{eq.a3}) that is {\it compatible} with controller design.
It is tractable for controller design and its implementation  that the model is linear to the input $u$.
To this end, we further specialize the class of the lifting function $\phi(z)$ in the following form 
\begin{align}
\phi(z)=\left[\begin{array}{c}
\psi(x)\\
u
\end{array}
\right]
\in \mathbb{R}^{N+m},
\nonumber
\end{align}
where $\psi(x):\mathbb{R}^{n} \to \mathbb{R}^{N}$ is $N$-dimensional lifting function given by
\begin{align}
\psi(x)=\left[\begin{array}{c}
\psi_{1}(x)\\
\vdots\\
\psi_{N}(x)
\end{array}
\right]
\in \mathbb{R}^{N},
\nonumber
\end{align}
and $N+m=N_{\phi}$ holds.
Let the matrix $\mathcal{A}$ be partitioned as 
 \begin{align}
\mathcal{A}=\left[\begin{array}{cc}
A & B\\
* & *
\end{array}
\right]
\in \mathbb{R}^{(N+m)\times(N+m)},
\nonumber
\end{align}
where $A\in \mathbb{R}^{N\times N}$ and $B \in \mathbb{R}^{N\times m}$.
Then, it follows from (\ref{eq.a3}) that the following expression of the time-evolution of $\psi(x)$ holds. 
\begin{align}
\psi(x(k+1))\approx A\psi(x(k))+Bu(k).
\nonumber
\end{align}

In addition, we give the approximation of the output equation in (\ref{eq.a0}) by
\begin{align*}
y(k) \approx C\psi(x(k)),
\end{align*}
where $C \in \mathbb{R}^{l\times N}$. For simplicity of notation, we let $\psi(k)=\psi(x(k))$. 
Then, we obtain the state-space model defined on the functional space as 
\begin{align}
	\left\{
	\begin{array}{ll}
	\psi(k+1)&=A\psi(k)+Bu(k),\\
	y(k)&=C\psi(k).
	\end{array}
	\right.
	\label{eq.a10}
\end{align}
The model (\ref{eq.a10}) approximately describes the nonlinear input-output behavior generated by (\ref{eq.a0}).
In this paper, the model (\ref{eq.a10}) is called the ``Koopman model''.
The aim of this paper is to propose the learning method of the system matrices $\left(A,B,C\right)$ based on some data-sets.



\subsubsection{{Learning Koopman Operator}}

For simplicity of notation, we define the following data-matrices based on the sequences of the input, output, and state of the system (\ref{eq.a0}).
\begin{align*}
	U_{k}&:=\left[u(k)~u(k+1)~\cdots~u(k+M-1)\right]\in \mathbb{R}^{m\times M},\\
	Y_{k}&:=\left[y(k)~y(k+1)~\cdots~y(k+M-1)\right]\in \mathbb{R}^{m\times M},\\
	\Psi_{k}&:=\left[\psi(k)~\psi(k+1)~\cdots~\psi(k+M-1)\right]\in \mathbb{R}^{N\times M},\\
	\Psi_{k+1}&:=\left[\psi(k+1)~ \psi(k+2)~\cdots~\psi(k+M)\right] \in \mathbb{R}^{N\times M}.
\end{align*}
It should be noted that $\Psi_k$ and $\Psi_{k+1}$ are constructed by using the measured data on the state, $\{x(k),\ldots,x(k+M)\}$. 
In this paper, it is assumed that the data-set $\left(U_{k},Y_{k},\Psi_{k},\Psi_{k+1}\right)$ is given and available 
for learning the Koopman operator that expresses (\ref{eq.a0}).
The problem of learning is formulated as follows.

Given the data-matrices $(U_{k},Y_k, \Psi_{k},\Psi_{k+1})$, solve the optimization problem:
\begin{align}
	 \min_{A,B,C} &\quad J_{1}(A,B) + J_2(C),
	\label{eq:ORopt}
\end{align}
where $J_{1}(A,B)$ and $J_2(C)$ are given by
 \begin{align}
J_{1}(A,B)
&:= \left\| 
\Psi_{k+1}
-\left[A~B\right] 
\left[\begin{array}{c}
\Psi_{k}\\ U_{k}
\end{array}
\right]
\right\|_{F}^2,
\label{eq.b1}\\
J_2(C) &:= \left\|Y_{k}-C\Psi_{k}\right\|_{F}^2.
\label{eq.a9.5}
\end{align}
The solution to the optimization problem (\ref{eq:ORopt}) provides the system matrices $(A^{\dagger},B^{\dagger},C^{\dagger})$ of the Koopman model (\ref{eq.a10}).
It is assumed that $\left[\Psi_k^{\top} U_k^{\top}\right]^{\top}$ is of full row rank, which is a natural assumption when rich data is available for learning.  Then, the learned matrices $(A^{\dagger},B^{\dagger},C^{\dagger})$ are uniquely determined by any given data.

\begin{rem}
\label{rem:ORopt}
Note that the optimization problem (\ref{eq:ORopt}) is {\it separately} solvable: 
the minimization of just $J_1(A,B)$ provides the optimal $(A^{\dagger},B^{\dagger})$,
while that of $J_2(C)$ provides the optimal $C^{\dagger}$.  
\end{rem}

\subsection{Dissipativity}

In this subsection, we review {\it dissipativity} of dynamical systems. 
Dissipativity is a property of characterizing dynamical systems and plays an important role in system analysis, in particular, the analysis of feedback or more general interconnection of dynamical systems (see e.g., the pioneering work by \cite{Willems1972} and developments e.g., by \cite{Hill_76,Hill_77}). 
Dissipativity is defined for the input-output system (\ref{eq.a0}) as follows.

\begin{defi}
Given a scalar function $s(u,y): \mathbb{R}^{m + l} \to \mathbb{R}$, 
the system (\ref{eq.a0}) is said to be {\it dissipative} for $s(u, y)$ if there is a non-negative function $V(x):\mathbb{R}^{n} \to \mathbb{R}_+$ such that the inequality
\begin{align}
	V(x(k))-V(x(0)) \leq \sum_{\tau=0}^{k} s(u(\tau),y(\tau)),\quad \forall k \geq 0
	\label{eq:DI}
\end{align}
holds.
\end{defi}

The functions $s(u,y)$ and $V(x)$ and the inequality (\ref{eq:DI}) are called 
the supply rate, storage function, and dissipation inequality, respectively.  
In the view of dissipativity theory, dynamical systems are briefly modeled by an {\it one-dimensional} difference {\it inequality} (\ref{eq:DI}), instead of a {\it multi-dimensional} difference {\it equation} (\ref{eq.a0}).

A characterization of dissipative {\it linear} dynamical systems is given. 
We specialize the supply rate $s(u,y)$ in the quadratic form as
\begin{align}
s(u,y)=
-\left[
\begin{array}{c}
y\\
u
\end{array}
\right]^{\mathrm{T}}
\Xi
\left[
\begin{array}{c}
y\\
u
\end{array}
\right], 
\label{eq.b0.6}
\end{align}
where $\Xi$ is the real symmetric matrix of
\begin{align}
\Xi
= \left[
\begin{array}{cc}
\Xi_{11} & \Xi_{12} \\ \Xi_{12}^{\mathrm{T}} & \Xi_{22}
\end{array}
\right].
\label{eq.b2}
\end{align}
Even for the specialization of the supply rate, the dissipativity includes some important property of dynamical systems.
For example, the dissipativity with respect to $(\Xi_{11}, \Xi_{12}, \Xi_{22}) = (0, -1, 0)$ represents the passivity of dynamical systems, 
and that with respect to $(\Xi_{11}, \Xi_{12}, \Xi_{22}) = (1, 0, -\gamma)$ for some positive constant $\gamma$ represents the bounded $L_2$ gain.

The dissipativity of linear input-output systems, e.g., described by  (\ref{eq.a10}), is characterized  by the following lemma.

\begin{lemm}
The following statements (i) and (ii) are equivalent (see the book by \cite{brogliato2007dissipative}).
\begin{description}
\item [(i)]~The Koopman model (\ref{eq.a10}) is dissipative for the supply rate $s(u,y)$ of (\ref{eq.b0.6}). 
\item [(i\hspace{-0.1em}i)]~There exists a symmetric matrix $P$ such that the following inequalities hold.
\begin{align}
&P \succ 0\label{eq.b3_1}\\
&\left[
\begin{array}{cc}
A & B\\I & 0
\end{array}
\right]^{\mathrm{T}}
\left[
\begin{array}{cc}
P & 0\\0 & -P
\end{array}
\right]
\left[
\begin{array}{cc}
A & B\\I & 0
\end{array}
\right]
+\Theta \prec 0,
\label{eq.b3}
\end{align}
where 
\begin{align*}
\Theta = 
\left[
\begin{array}{cc}
C & 0\\ 0 & I
\end{array}
\right]^{\mathrm{T}}
\Xi
\left[
\begin{array}{cc}
C & 0\\ 0 & I
\end{array}
\right].
\end{align*}
\end{description}
\end{lemm}

\section{Learning Koopman Operator with Dissipativity-Constraints}
\label{sec3}

In this section, we propose a learning method of nonlinear dynamical systems with incorporating {\it a priori} information on the system dissipativity.
We assume that the supply rate $s(u,y)$ characterizing the system dissipativity is already given and available for learning.  
Then, we aim at incorporating the dissipativity information into the Koopman model (\ref{eq.a10}).


\subsection{Problem Setting} 

We aim at constructing the Koopman model (\ref{eq.a10}) that satisfies the dissipation inequality (\ref{eq:DI}) based on some data-sets.
This learning problem of the system matrices $\left(A,B,C\right)$ of (\ref{eq.a10}) is reduced to the problem 
of (\ref{eq:ORopt}) subject to the dissipativity constraints (\ref{eq.b3_1}) and (\ref{eq.b3}).
The problem is mathematically formulated as follows.
\smallskip

\begin{problem}
\label{prob:1}
Given the real symmetric matrix $\Xi$ and the data-matrices $\left(U_{k},Y_k,\Psi_{k},\Psi_{k+1}\right)$, solve the optimization problem: 
\begin{align*}
	\min_{P,A,B,C}&\quad {J_{1}(A,B)} + J_2(C)\\
	\mbox{sub to}&\quad(\ref{eq.b3_1}), (\ref{eq.b3}).
\end{align*}
\end{problem}
\smallskip

Suppose that the optimal solution to Problem \ref{prob:1} is given by $(P^{*},A^{*},B^{*},C^{*})$.
Then the Koopman model (\ref{eq.a10}) with $(A^{*},B^{*},C^{*})$ is dissipative for the supply rate $s(u,y)$ of (\ref{eq.b0.6}). 

Note that the dissipativity constraint, described by (\ref{eq.b3_1}) and (\ref{eq.b3}), is non-convex in decision variables  $\left(P,A,B,C\right)$. 
It is not numerically tractable to solve Problem \ref{prob:1}.
In the next subsection, we try to approximately reduce the problem to a convex one.

\subsection{Convex Approximation of Problem \ref{prob:1}} 

On the basis of the variable transformation technique by \cite{hoagg2004first,abe2016subspace},  
the nonlinear inequality (\ref{eq.b3}) is reduced into a linear matrix one. 
First, we expand the inequality (\ref{eq.b3}) as
\begin{align}
\left[
\begin{array}{cc}
P-C^{\mathrm{T}}\Xi_{11}C & -C^{\mathrm{T}}\Xi_{12}\\ 
-\Xi_{12}^{\mathrm{T}} C & -\Xi_{22}
\end{array}
\right]
-\left[A~B\right]^{\mathrm{T}}P\left[A~B\right] \succ 0.
\label{eq.b5}
\end{align}
Noting $P \succ 0$ and applying the Schur complement to (\ref{eq.b5}), it follows that
\begin{align}
\left[
\begin{array}{cc}
\left[\begin{array}{cc}
P-C^{\mathrm{T}}\Xi_{11}C & -C^{\mathrm{T}}\Xi_{12}\\ 
-\Xi_{12}^{\mathrm{T}} C & -\Xi_{22}
\end{array}
\right] & 
\left[
\begin{array}{c}
A^{\mathrm{T}} P\\
B^{\mathrm{T}} P
\end{array}
\right]\\
\left[PA~PB\right] & P
\end{array}
\right]\succ 0
\label{eq.b10}
\end{align}
holds.

Next, we apply the variable transformation to $(P, A, B)$.  We let
\begin{align}
R=PA,\quad S=PB
\label{eq.b10.1}
\end{align}
to reduce (\ref{eq.b10}) to the inequality 
\begin{align}
\left[
\begin{array}{ccc}
P-C^{\mathrm{T}}\Xi_{11}C & -C^{\mathrm{T}}\Xi_{12} &  R^{\mathrm{T}}  \\ 
-\Xi_{12}^{\mathrm{T}} C  & -\Xi_{22} &S^{\mathrm{T}}\\
  R & S & P
\end{array}
\right]\succ 0.
\label{eq.b11}
\end{align}
Note that this (\ref{eq.b11}) is still equivalent to (\ref{eq.b3}).
This is because that the solution $(P,R,S,C)$ to (\ref{eq.b11}) generates 
the solution to (\ref{eq.b3}) by $(P,A,B,C) = (P,P^{-1}R,P^{-1}S,C)$.

Now, suppose that $C$ is given, e.g., by just minimizing $J_2(C)$ based on the data-set $(Y_k,\Phi_k)$.
Then, the inequality (\ref{eq.b11}) is linear in the matrices $(P,R,S)$, which is numerically tractable.

There is the drawback in the variable transformation of (\ref{eq.b10.1}):
the cost function  $J_{1}(A,B)$ of (\ref{eq.b1}) becomes non-convex in the transformed variables $(P,R,S)$, which is numerically intractable. 
To overcome the drawback and to numerically obtain the feasible solution to Problem \ref{prob:1}, 
we approximately transform $J_{1}(A,B)$ to a convex one.
To this end, we introduce $W =P$, as a weighting matrix, into $J_{1}(A,B)$ to define a new cost function as follows.
\begin{align}
J_{1, W}(P,R,S)
&= \left\| W 
\left(
\Psi_{k+1}-\left[A~B\right] \left[
\begin{array}{c}
\Psi_{k}\\ U_{k}
\end{array}
\right]
\right)
\right \|_{F}^2\nonumber\\
&= \left\| 
P \Psi_{k+1}-\left[R~S\right] \left[\begin{array}{c}
\Psi_{k}\\ U_{k}
\end{array}
\right]
\right\|_{F}^2.
\label{eq.b12}
\end{align}
The function $J_{1, W}(P,R,S)$ of (\ref{eq.b12}) is convex in the matrices $(P,R,S)$.
The minimization problem of  $J_{1, W}(P,R,S)$ under the inequalities (\ref{eq.b3_1}) and (\ref{eq.b11}) is in the class of the  convex optimization.
The  optimization problem is summarized as follows.
\smallskip

\begin{problem}
\label{prob:2}
Given the system matrix $C$, the real symmetric matrix $\Xi$, and the data-matrices $\left(U_{k},\Psi_{k},\Psi_{k+1}\right)$, solve the optimization problem: 
\begin{align*}
\min_{P,R,S}&\quad{J_{1, W}(P,R,S)}\\
\mathrm{sub\ to}&\quad(\ref{eq.b3_1}), (\ref{eq.b11}).
\end{align*}
\end{problem}
\smallskip

Suppose that Problem \ref{prob:2} is feasible and that  the solution $(\hat{P},\hat{R},\hat{S})$ is given. 
Then, we obtain the system matrices as
\begin{align}
\hat{A}=\hat{P}^{-1} \hat{R},
\quad \hat{B}=\hat{P}^{-1} \hat{S}.
\label{eq:sysmat}
\end{align}

We have the following proposition for the constructed model based on the solution to Problem \ref{prob:2}.

\begin{prop}
Suppose that Problem \ref{prob:2} is feasible and that  system matrices are given by (\ref{eq:sysmat}).  
Then, the quadruplet $(\hat{P},\hat{A},\hat{B},C)$ is the {\it feasible} solution to Problem \ref{prob:1}. 
In other words, the Koopman model (\ref{eq.a10}) with $(\hat{A},\hat{B},C)$ is dissipative for the supply rate $s(u,y)$ of (\ref{eq.b0.6}). 
\end{prop}

As implied in the proposition, the solution to Problem \ref{prob:2} is the {\it feasible} solution, 
but may not be the {\it optimal} solution to Problem \ref{prob:1}.
In general, the solution $(\hat{P},\hat{A},\hat{B},C)$ is conservative for Problem \ref{prob:1}.
In the following subsection, we aim at finding a better approximation of Problem \ref{prob:1}.

\subsection{Sequential Convex Approximation of Problem \ref{prob:1}}

In this subsection, we give the efficient solution method for Problem \ref{prob:1} based on the overbounding method proposed by \cite{sebe2018sequential}.
In the overbounding method, the inner approximations of nonlinear matrix inequalities are sequentially constructed.
This sequential method contributes to gradually reduce the conservativeness of the solution of Problem \ref{prob:2}.

Suppose that Problem \ref{prob:2} is feasible and that the feasible solution to Problem \ref{prob:1}, denoted by $(\hat{P},\hat{A},\hat{B})$, is constructed.
Then, we try to update the {\it initial guess}  $(\hat{P},\hat{A},\hat{B})$ to reduce the conservativeness, i.e., to further reduce $J_1(A,B)$.
First, we transform the decision variables of Problem \ref{prob:1}, denoted by $(P,A,B)$, into $(\Delta P,\Delta A,\Delta B)$ as follows.
\begin{align*}
P = \hat{P}+\Delta P,\quad A = \hat{A}+\Delta A,\quad B = \hat{B}+\Delta B.
\end{align*}
Further, we let $G$ and $H$ be additional decision variables.
With those $G$ and $H$, we define the inequality condition described in (\ref{eq.c8}), where 
$Q(\Delta P, \Delta A, \Delta B)$ is given by (\ref{eq.c4}) and $F(\Delta P)$ is given by
\begin{figure*}
\begin{align}
\mathrm{He}\left(
\left[
\begin{array}{ccc}
Q(\Delta P,\Delta A,\Delta B) & \left[
\begin{array}{c}
0 \\ \Delta P 
\end{array}
\right] & 0 \\
0  & -G & G\\ 
-H \left[
\begin{array}{ccc}
\Delta A & \Delta B & 0
\end{array}
\right] & 0 & -H
\end{array}
\right]
\right)\prec 0.
\label{eq.c8}
\end{align}
\end{figure*}
\begin{figure*}
\begin{align}
Q(\Delta P,\Delta A,\Delta B) &= -\cfrac{1}{2}\left[
\begin{array}{cc}
-F(\Delta P) & 0\\0 & \hat{P}+\Delta P
\end{array}
\right]
+
\left[
\begin{array}{ccc}
0 & 0 & 0\\ -\hat{P}\hat{A} & -\hat{P}\hat{B} & 0
\end{array}
\right]\nonumber\\
&\hspace{5mm}+
\left[
\begin{array}{ccc}
0 & 0 & 0\\ -\Delta P\hat{A} & -\Delta P\hat{B} & 0
\end{array}
\right]
+
\left[
\begin{array}{ccc}
0 & 0 & 0\\ -\hat{P}\Delta A & -\hat{P}\Delta B & 0
\end{array}
\right]
\label{eq.c4}.
\end{align}
\end{figure*}
\begin{align*}
F(\Delta P)=\left[
\begin{array}{cc}
\hat{P}+\Delta P -C^{\mathrm{T}}\Xi_{11}C & -C^{\mathrm{T}}\Xi_{12}\\ 
-\Xi_{12}^{\mathrm{T}} C & -\Xi_{22}
\end{array}
\right].
\end{align*}
We show that (\ref{eq.c8}) is a sufficient condition for (\ref{eq.b3}) as stated in the following proposition. 

\begin{prop}
\label{prop:3}
Suppose (\ref{eq.c8}) holds. 
Then, letting $(P, A, B) =( \hat{P} + \Delta P,  \hat{A} + \Delta A,  \hat{B} + \Delta B)$, it holds that (\ref{eq.b3}).
\end{prop}

The proof follows Proposition \ref{prop:3} in the work by \cite{sebe2018sequential} and is omitted in this paper.
Furthermore, it should be noted that (\ref{eq.c8}) is linear in $(\Delta P,\Delta A,\Delta B, G)$.
This implies that 
for any fixed $H$, (\ref{eq.c8}) is in the form of LMIs and is numerically tractable.

Recall $J_{1}(A,B)$ of $(\ref{eq.b1})$ to obtain the expression 
\begin{align}
&J_{1}(\hat{A}+\Delta A,\hat{B}+\Delta B)\nonumber\\
 &=\left\| \Psi_{k+1}-\left[\hat{A}+\Delta A,~\hat{B}+\Delta B \right] \left[\begin{array}{c}
\Psi_{k}\\ U_{k}
\end{array}
\right]
\right\|_{F}^2.
\label{eq.c10}
\end{align}
Then, the problem of  finding 
$(\Delta P,\Delta A,\Delta B, G)$ that minimizes $J_{1}(\hat{A}+\Delta A,\hat{B}+\Delta B)$ 
under the constraint (\ref{eq.c8}) 
 based on  the initial guess  $(\hat{P},\hat{A},\hat{B})$ is stated as follows
\smallskip

\begin{problem}
\label{prob:3}
Given the system matrix $C$, the real symmetric matrix $\Xi$, the data-matrices $\left(U_{k},\Psi_{k},\Psi_{k+1}\right)$, 
the feasible solution to Problem \ref{prob:1}, denoted by $(\hat{P},\hat{A},\hat{B})$, and the real matrix $H$,
solve the optimization problem:
\begin{align*}
	\min_{\Delta P,\Delta A,\Delta B,G}&\quad J_{1}(\hat{A}+\Delta A,\hat{B}+\Delta B)\\
	\mathrm{sub\ to}& \quad(\ref{eq.c8}),~\hat{P}+\Delta P\succ 0.
\end{align*}
\end{problem}

\smallskip

With the optimal solution $(\Delta \bar{P}, \Delta \bar{A}, \Delta \bar{B})$  to Problem \ref{prob:3}, we obtain the matrices 
\begin{align*}
\bar{P} = \hat{P}+\Delta \bar{P},\quad \bar{A} = \hat{A}+\Delta \bar{A},\quad \bar{B} = \hat{B}+\Delta \bar{B}.
\end{align*}

The notation on the optimal solutions to Problems  
\ref{prob:1}--\ref{prob:3} are summarized on Table \ref{tb1}.

\begin{table}[t]
\begin{center}
\caption{Notation of optimal solutions}
\label{tb1}
\begin{tabular}{cc}
Problem & Solution \\\hline
Unconstrained (\ref{eq:ORopt})  &   $(A^{\dagger},B^{\dagger},C^{\dagger})$\\
Problem \ref{prob:1} & $(P^{*},A^{*},B^{*},C^{*})$  \\
Problem \ref{prob:2} & $(\hat{P},\hat{R},\hat{S}) \to (\hat{P},\hat{A},\hat{B})$  \\
Problem \ref{prob:3} &  $(\Delta \bar{P}, \Delta \bar{A}, \Delta \bar{B}) \to (\bar{P},\bar{A},\bar{B})$ \\ \hline
\end{tabular}
\end{center}
\end{table}

Note that the solution $(\bar{P},\bar{A},\bar{B})$  is the less conservative solution to Problem \ref{prob:1}  than the initial guess $(\hat{P},\hat{A},\hat{B})$ for any real matrix $H$ satisfying
\begin{align}
H+H^{\mathrm{T}} \succ 0.
\label{H-condition}
\end{align}
This fact is mathematically stated in the following proposition.  

\begin{prop}
\label{prop:4}
Suppose that Problem \ref{prob:2} is feasible.
Then, for any real matrix $H$ satisfying the condition (\ref{H-condition}), Problem \ref{prob:3} is feasible.
In addition, if  $(\hat{P},\hat{A},\hat{B})$ is not the solution to Problem \ref{prob:1}, the strict inequality
\begin{align}
	&J_{1}(\bar{A}, \bar{B}) < J_{1}(\hat{A},\hat{B}) 
\label{cost}
\end{align}
holds.
\end{prop}

{\bf Proof.}
First, we prove that Problem \ref{prob:3} has at least one feasible solution $(\Delta P,\Delta A,\Delta B,G)=(\Delta P_{f},\Delta A_{f},\Delta B_{f},G_{f})$.
Let $(\Delta P_{f},\Delta A_{f},\Delta B_{f})=\left(0,0,0\right)$.
Then, the left hand side of (\ref{eq.c8}) is reduced to 
\begin{align}
\mathrm{He}\left(
\left[
\begin{array}{cc}
Q(0,0,0) & 0 \\
0  & X
\end{array}
\right]
\right),
\label{eq.pro1}
\end{align}
where
\begin{align}
Q(0,0,0)&=-\frac{1}{2}
\left[
\begin{array}{cc}
\left[\begin{array}{cc}
\hat{P}-C^{\mathrm{T}}\Xi_{11}C & -C^{\mathrm{T}}\Xi_{12}\\ 
-\Xi_{12}^{\mathrm{T}} C & -\Xi_{22}
\end{array}
\right] & 
\left[
\begin{array}{c}
\hat{A}^{\mathrm{T}} \hat{P}\\
\hat{B}^{\mathrm{T}} \hat{P}
\end{array}
\right]\\
\left[\hat{P}\hat{A}~\hat{P}\hat{B}\right] & \hat{P}
\end{array}
\right],
\nonumber\\
X&=
\left[
\begin{array}{cc}
-G & G\\ 
0 & -H
\end{array}
\right].
\nonumber
\end{align}
Since $(\hat{P},\hat{A},\hat{B})$ satisfies the inequality (\ref{eq.b11}), we see that
\begin{align}
\mathrm{He}\left(Q(0,0,0)\right) \prec 0
\label{eq.pro2}
\end{align}
holds.
In addition, letting $G_{f}=\frac{1}{2}\left(H+H^{\mathrm{T}}\right)$, we have
\begin{align}
\mathrm{He}\left(X\right)=-\left[\begin{array}{cc}
(H+H^{\mathrm{T}}) & -\frac{1}{2}(H+H^{\mathrm{T}}) \\
-\frac{1}{2}(H+H^{\mathrm{T}}) & (H+H^{\mathrm{T}})
\end{array}
\right]\prec 0.
\label{eq.pro3}
\end{align}
It follows from (\ref{eq.pro2}) and (\ref{eq.pro3}) that
\begin{align}
(\Delta P_{f},\Delta A_{f},\Delta B_{f},G_{f})=\left(0,0,0,\frac{1}{2}(H+H^{\mathrm{T}})\right)
\label{eq.pro4}
\end{align}
satisfies the constraint (\ref{eq.c8})  and is the feasible solution to Problem \ref{prob:3}. 

Next, we prove that (\ref{cost}) holds.
Since the constraint (\ref{eq.c8}) is the strict inequality, i.e., the feasible region is open, the neighborhood of (\ref{eq.pro4}) still gives the feasible solution.
In addition, recall that the cost function $J_{1}(\hat{A}+\Delta A,\hat{B}+\Delta B)$ is strongly convex in $(\Delta A, \Delta B)$. 
Then, $J_{1}(\hat{A}+\Delta A,\hat{B}+\Delta B)$ strictly decreases for some point in the neighborhood compared with $J_{1}(\hat{A}+\Delta A_{f},\hat{B}+\Delta B_{f})$. 
This completes the proof of the proposition. 
$\Box$

On the basis of the fact stated in Proposition \ref{prop:4}, a sequential algorithm of solving Problem \ref{prob:1} is proposed. 
Suppose that  Problem \ref{prob:3} with $(\hat{P},\hat{A},\hat{B}, H) = (\bar{P}_i,\bar{A}_i,\bar{B}_i, H_i)$ has the optimal solution $(\Delta \bar{P}_i, \Delta \bar{A}_i, \Delta \bar{B}_i, \bar{G}_i)$.  
Consider the updating law
\begin{align}
	&(\bar{P}_{i+1},\bar{A}_{i+1},\bar{B}_{i+1}, H_{i+1})\nonumber\\ 
	&\leftarrow (\bar{P}_i+\Delta \bar{P}_i, \bar{A}_i+\Delta \bar{A}_i,\bar{B}_i+\Delta \bar{B}_i, \bar{G}_i).
	\label{eq:update}
\end{align}
Then, by sequentially solving Problem \ref{prob:3} with the updated $(\hat{P},\hat{A},\hat{B}, H) = (\bar{P}_{i+1},\bar{A}_{i+1},\bar{B}_{i+1}, H_{i+1})$, 
we obtain the solution $(\Delta \bar{P}_{i+1}, \Delta \bar{A}_{i+1}, \Delta \bar{B}_{i+1}, \bar{G}_{i+1})$, which generates the less conservative solution to Problem \ref{prob:1}.  

We propose to sequentially solve Problem \ref{prob:3} with updated $(\bar{P}_{i+1},\bar{A}_{i+1},\bar{B}_{i+1}, H_{i+1})$ 
to obtain the less conservative solution to Problem \ref{prob:1}.
The sequential solution method is summarized in Algorithm.


\begin{itembox}[l]{{\it Algorithm}}
\begin{itemize}
\item[Step0] 
Find the solution $(\hat{P},\hat{R},\hat{S})=(\hat{P}_0,\hat{R}_0,\hat{S}_0)$ to Problem \ref{prob:2} and obtain 
$\bar{A}_{0}=\hat{P}_0^{-1} \hat{R}_0$ and $\bar{B}_{0}= \hat{P}_0^{-1} \hat{S}_0$.  In addition, let $H_{0}=I$ and $i = 0$.
\item[Step1]
Given $(\hat{P},\hat{A},\hat{B}, H) = (\bar{P}_i,\bar{A}_i,\bar{B}_i, H_i)$, 
find the solution $(\Delta \bar{P}_i, \Delta \bar{A}_i, \Delta \bar{B}_i, \bar{G}_i)$ to Problem \ref{prob:3}.
\item[Step2]
Apply (\ref{eq:update}) to obtain $(\bar{P}_{i+1},\bar{A}_{i+1},\bar{B}_{i+1}, H_{i+1})$.  Then,
set $i \leftarrow i+1$ and go to Step1.
\end{itemize}
\end{itembox}

\begin{figure*}[t]
\centering
\includegraphics[scale=0.5]{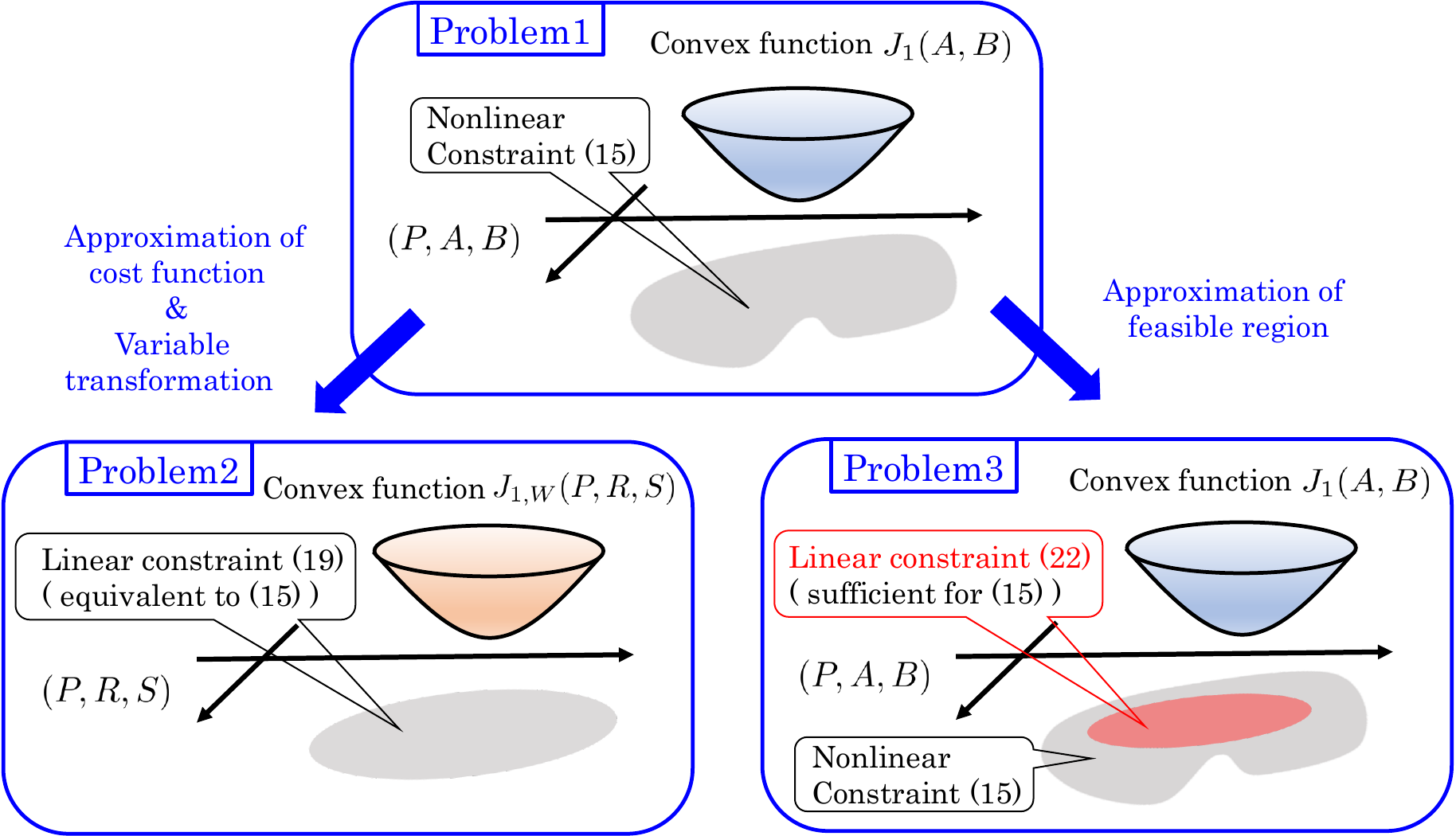}
\caption{Cost functions and feasible regions of Problems  
\ref{prob:1}--\ref{prob:3}}
\label{fig1}
\end{figure*}

Finally, the relationship among Problems \ref{prob:1}--\ref{prob:3} is illustrated in \rfig{fig1}.
Recall that Problem \ref{prob:1} is the minimization problem of the convex cost function $J_1$ under the nonlinear matrix inequality (\ref{eq.b3}).
The problem is approximated in different two ways as Problems \ref{prob:2} and \ref{prob:3}. 
In Problem \ref{prob:2}, $J_1$ is distorted and replaced by another convex one $J_{1,W}$, while (\ref{eq.b3}) is equivalently reduced to the LMI (\ref{eq.b11}).
In Problem \ref{prob:3}, the inner convex approximation of  (\ref{eq.b3}) is found and utilized, while keeping $J_1$.
Problem \ref{prob:2} is solved aiming at finding the feasible solution to Problem \ref{prob:1}, while 
Problem \ref{prob:3} is sequentially solved aiming at reducing the conservativeness of the solution. 


\section{Numerical Experiment}
\label{sec4}
In this section, we demonstrate the procedure of learning a nonlinear dynamical system by applying the proposed algorithm.
Consider a continuous-time nonlinear dynamical system described by
\begin{align}
\left\{
\begin{array}{lll}
\dot{x}_1(t) &= x_2(t),\\
\dot{x}_2(t) &= -2x_2(t)+x(1)\cos\left(x_1(t)+x_2(t)\right)+u(t),\\
y(t)&= x_2(t).
\end{array}
\right.
\label{eq.e1}
\end{align}
It is known that the system is dissipative with respect to $(\Xi_{11}, \Xi_{12}, \Xi_{22}) = (0, -1, 0)$, 
i.e., the system is passive (see e.g., the work by \cite{zakeri2019passivity}).
In this experiment, we aim at accurately learning the dynamical system in the Koopman model (\ref{eq.a10}), 
while incorporating the passivity property.

In the experimental setup, we consider that the time series of $x(t)$, $u(t)$, and $y(t)$ are sampled
at each 0.01 time period from the system (\ref{eq.e1}), which are denoted by  $\{x(k)\}$, $\{u(k)\}$, and $\{y(k)\}$, respectively.
The input series $\{u(k)\}$ for learning are determined from randomly  selected values from the uniform distribution  in $[\,-1,\,1\,]$.
Then, the state and output series $\{x(k)\}$ and $\{y(k)\}$ are measured synchronously.
In total, the data at 5000 samples are obtained.

We try to apply Algorithm to the data  $\{x(k)\}$, $\{u(k)\}$, and $\{y(k)\}$.
To this end, first, we let $(\Xi_{11}, \Xi_{12}, \Xi_{22}) = (0, -1, -0.2)$ and its corresponding dissipativity constraint be defined 
to inherit the {\it a priori} information on the passivity in a relaxed form.
Furthermore, let the lifting function $\psi(x(k))$ be composed of the state $x(k)=[\,x_{1}(k),x_{2}(k)\,]^{\mathrm{T}}$ 
and thin plate spline radial basis functions $\psi_i(x(k))$, $i \in \{1, 2,\ldots ,8\}$,
where $\psi_i(x(k))$ is given by
\begin{align*}
\psi_i(x(k)) = \|x(k)-r_i\|_2^2\ln{\|x(k)-r_i\|_2}
\end{align*}
and  the values of $r_i$ are selected randomly from the uniform distribution on the unit box.
Then, the lifting function $\psi(x(k))$ is described by
\begin{align}
\psi(x(k))=\left[\,x(k)\ \psi_1(x(k))\ \cdots\ \psi_{8}(x(k))\,\right]^\mathrm{T}
\in \mathbb{R}^{2+8}.
\label{eq.e2}
\end{align}

By applying  Algorithm, we constructed the dissipativity-constrained Koopman model, which approximates the nonlinear system (\ref{eq.e1}) and is called Model 1.
In addition, we constructed two different models by using the same time series data: 
(Model 2) one is the no-constrained Koopman model, which is simply constructed by solving (\ref{eq:ORopt}), 
while (Model 3) the other is the dissipativity-constrained {\it linear} model, which is based on $\psi(x(k)) = x(k)$ and is constructed by applying the learning method proposed by \cite{abe2016subspace}.

\begin{figure}[th]
\centering
\includegraphics[scale=0.8]{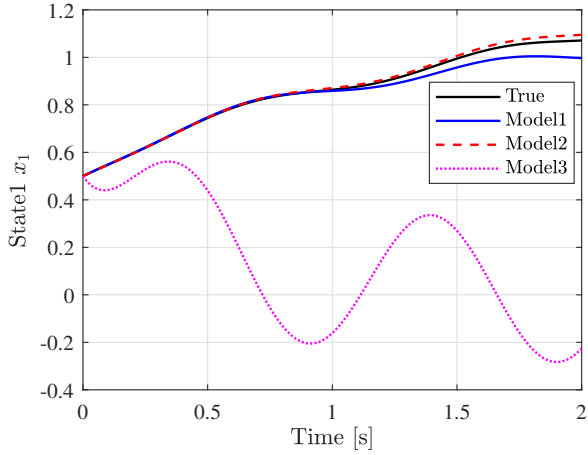}
\caption{State trajectory $x_1(k)$.}
\label{state1}
\end{figure}

\begin{figure}[th]
\centering
\includegraphics[scale=0.8]{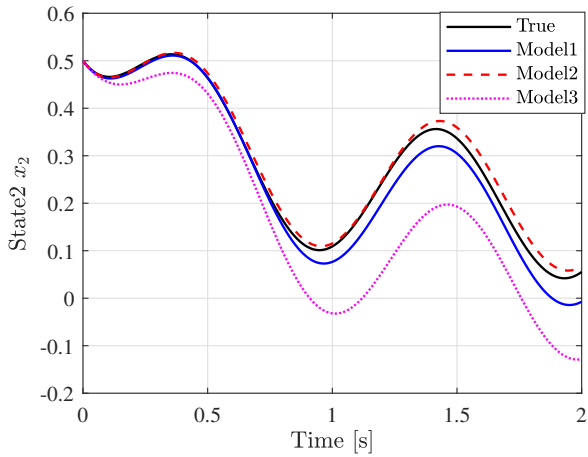}
\caption{State trajectory $x_2(k)$.}
\label{state2}
\end{figure}

First, to show the model accuracy, the three models are compared with the true nonlinear system (\ref{eq.e1}).
The result of the frequency response against the sin-wave input is illustrated in \rfigs{state1} and \ref{state2}, 
where the state trajectory of the models is shown.
In the figures, the black solid, blue solid, red dashed, and pink dotted lines represent 
the state of the true system, Model 1 (proposed model), Model 2, and Model 3, respectively.
We see from Fig.~\ref{state1} that Models 1 and 2, i.e., the Koopman models, accurately express the nonlinear behavior generated by (\ref{eq.e1}), 
while Model 3, i.e., the linear model, is not.  The lifting function with the basis (\ref{eq.e2}) contributes to improving the ability of model expression.


%

\begin{figure}[th]
\centering
\includegraphics[scale=0.8]{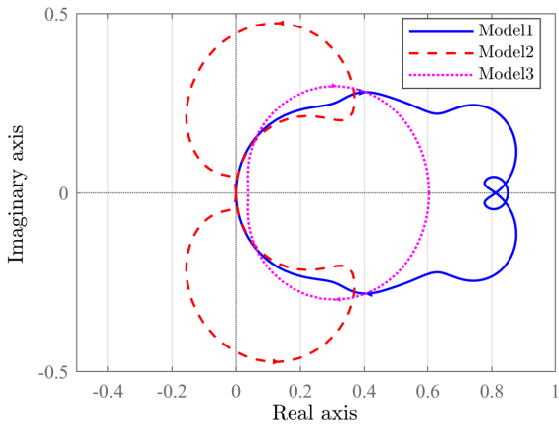}
\caption{Comparison of Nyquist plot}
\label{nyquist}
\end{figure}

Next, to show the validity of the dissipativity constraint, we define the transfer function for the Koopman model.
Letting $G(e^{j\omega T})=C\left(e^{j\omega T}I-A\right)^{-1}B$, we reduce the dissipativity constraint, characterized by $(\Xi_{11}, \Xi_{12}, \Xi_{22}) = (0, -1, -0.2)$, to
\begin{align*}
	\mathrm{Re}[G(e^{j\omega T})]\geq -0.1,\quad \forall \omega\in\mathbb{R}.
\end{align*}
Then, the Nyquist plot of the $G(e^{j\omega T})$ for the three models is illustrated in \rfig{nyquist}.
As illustrated in the figure, Models 1 and 3 satisfy the dissipativity constraint, while Model 2 violates.
This shows that the dissipativity constraint imposed on the learning problems is valid.

\section{Conclusion}
\label{sec5}

This paper addressed the learning problem of nonlinear dynamical systems with incorporating the {\it a priori} information on the quadratic dissipativity.  
The problem was reduced to the data-driven approximation of the Koopman operator under the dissipativity constraint, which was called Problem  \ref{prob:1} in this paper.  
Then, the solution method to the problem was given and summarized in Algorithm.  
There are two main contributions of this paper. 
1) One is in this numerically efficient algorithm, which sequentially solves LMIs. 
2) The other is the performance analysis of the algorithm, which is stated in Proposition \ref{prop:4}.
In the analysis, it is guaranteed that the Koopman model constructed by the algorithm fits given data more accurately than the model defined at any initial guess.

\bibliography{ifacconf}             

\begin{thebibliography}{22}
\providecommand{\natexlab}[1]{#1}
\providecommand{\url}[1]{\texttt{#1}}
\providecommand{\urlprefix}{URL }
\expandafter\ifx\csname urlstyle\endcsname\relax
  \providecommand{\doi}[1]{doi:\discretionary{}{}{}#1}\else
  \providecommand{\doi}{doi:\discretionary{}{}{}\begingroup
  \urlstyle{rm}\Url}\fi

\bibitem[{Abe et~al.(2016)Abe, Inoue, and Adachi}]{abe2016subspace}
Abe, Y., Inoue, M., and Adachi, S. (2016).
\newblock Subspace identification method incorporated with a priori information
  characterized in frequency domain.
\newblock In \emph{2016 European Control Conference (ECC)}, 1377--1382. IEEE.

\bibitem[{Alenany et~al.(2011)Alenany, Shang, Soliman, and
  Ziedan}]{alenany2011improved}
Alenany, A., Shang, H., Soliman, M., and Ziedan, I. (2011).
\newblock Improved subspace identification with prior information using
  constrained least squares.
\newblock \emph{IET Control Theory \& Applications}, 5(13), 1568--1576.

\bibitem[{Brogliato et~al.(2007)Brogliato, Lozano, Maschke, and
  Egeland}]{brogliato2007dissipative}
Brogliato, B., Lozano, R., Maschke, B., and Egeland, O. (2007).
\newblock Dissipative systems analysis and control.
\newblock \emph{Theory and Applications}, 2.

\bibitem[{De~la Rosa and Yu(2016)}]{de2016randomized}
De~la Rosa, E. and Yu, W. (2016).
\newblock Randomized algorithms for nonlinear system identification with deep
  learning modification.
\newblock \emph{Information Sciences}, 364, 197--212.

\bibitem[{Goethals et~al.(2003)Goethals, Van~Gestel, Suykens, Van~Dooren, and
  De~Moor}]{goethals2003identification}
Goethals, I., Van~Gestel, T., Suykens, J., Van~Dooren, P., and De~Moor, B.
  (2003).
\newblock Identification of positive real models in subspace identification by
  using regularization.
\newblock \emph{IEEE Transactions on Automatic Control}, 48(10), 1843--1847.

\bibitem[{Hill and Moylan(1976)}]{Hill_76}
Hill, D. and Moylan, P. (1976).
\newblock The stability of nonlinear dissipative systems.
\newblock \emph{IEEE Transactions on Automatic Control}, 21(5), 708--711.

\bibitem[{Hill and Moylan(1977)}]{Hill_77}
Hill, D.J. and Moylan, P.J. (1977).
\newblock Stability results for nonlinear feedback systems.
\newblock \emph{Automatica}, 13(4), 377--382.

\bibitem[{Hoagg et~al.(2004)Hoagg, Lacy, Erwin, and Bernstein}]{hoagg2004first}
Hoagg, J.B., Lacy, S.L., Erwin, R.S., and Bernstein, D.S. (2004).
\newblock First-order-hold sampling of positive real systems and subspace
  identification of positive real models.
\newblock In \emph{Proceedings of the 2004 American control conference},
  volume~1, 861--866. IEEE.

\bibitem[{Inoue(2019)}]{inoue2019subspace}
Inoue, M. (2019).
\newblock Subspace identification with moment matching.
\newblock \emph{Automatica}, 99, 22--32.

\bibitem[{Jin et~al.(2016)Jin, Shao, Zhang, An, and Malekian}]{jin2016modeling}
Jin, X., Shao, J., Zhang, X., An, W., and Malekian, R. (2016).
\newblock Modeling of nonlinear system based on deep learning framework.
\newblock \emph{Nonlinear Dynamics}, 84(3), 1327--1340.

\bibitem[{Kevrekidis et~al.(2015)Kevrekidis, Rowley, and
  Williams}]{kevrekidis2015kernel}
Kevrekidis, I., Rowley, C., and Williams, M. (2015).
\newblock A kernel-based method for data-driven {K}oopman spectral analysis.
\newblock \emph{Journal of Computational Dynamics}, 2(2), 247--265.

\bibitem[{Koopman(1931)}]{koopman1931hamiltonian}
Koopman, B.O. (1931).
\newblock Hamiltonian systems and transformation in hilbert space.
\newblock \emph{Proceedings of the National Academy of Sciences of the United
  States of America}, 17(5), 315--318.

\bibitem[{Korda and Mezi{\'c}(2018)}]{korda2018linear}
Korda, M. and Mezi{\'c}, I. (2018).
\newblock Linear predictors for nonlinear dynamical systems: Koopman operator
  meets model predictive control.
\newblock \emph{Automatica}, 93, 149--160.

\bibitem[{Lacy and Bernstein(2003)}]{lacy2003subspace}
Lacy, S.L. and Bernstein, D.S. (2003).
\newblock Subspace identification with guaranteed stability using constrained
  optimization.
\newblock \emph{IEEE Transactions on automatic control}, 48(7), 1259--1263.

\bibitem[{Miller and De~Callafon(2013)}]{miller2013subspace}
Miller, D.N. and De~Callafon, R.A. (2013).
\newblock Subspace identification with eigenvalue constraints.
\newblock \emph{Automatica}, 49(8), 2468--2473.

\bibitem[{Okada and Sugie(1996)}]{okada1996subspace}
Okada, M. and Sugie, T. (1996).
\newblock Subspace system identification considering both noise attenuation and
  use of prior knowledge.
\newblock In \emph{Proceedings of 35th IEEE Conference on Decision and
  Control}, volume~4, 3662--3667. IEEE.

\bibitem[{Sebe(2018)}]{sebe2018sequential}
Sebe, N. (2018).
\newblock Sequential convex overbounding approximation method for bilinear
  matrix inequality problems.
\newblock \emph{IFAC-PapersOnLine}, 51(25), 102--109.

\bibitem[{Sitompul(2013)}]{sitompul2013adaptive}
Sitompul, E. (2013).
\newblock Adaptive neural networks for nonlinear dynamic systems
  identification.
\newblock In \emph{2013 Fifth International Conference on Computational
  Intelligence, Modelling and Simulation}, 8--13. IEEE.

\bibitem[{Willems(1972)}]{Willems1972}
Willems, J.C. (1972).
\newblock Dissipative dynamical systems part i: General theory.
\newblock \emph{Archive for Rational Mechanics and Analysis}, 45(5), 321--351.
\newblock \doi{10.1007/BF00276493}.

\bibitem[{Williams et~al.(2015)Williams, Kevrekidis, and
  Rowley}]{williams2015data}
Williams, M.O., Kevrekidis, I.G., and Rowley, C.W. (2015).
\newblock A data--driven approximation of the {K}oopman operator: Extending
  dynamic mode decomposition.
\newblock \emph{Journal of Nonlinear Science}, 25(6), 1307--1346.

\bibitem[{Yoshimura et~al.(2019)Yoshimura, Matsubayashi, and
  Inoue}]{yoshimura2019system}
Yoshimura, S., Matsubayashi, A., and Inoue, M. (2019).
\newblock System identification method inheriting steady-state characteristics
  of existing model.
\newblock \emph{International Journal of Control}, 92(11), 2701--2711.

\bibitem[{Zakeri and Antsaklis(2019)}]{zakeri2019passivity}
Zakeri, H. and Antsaklis, P.J. (2019).
\newblock Passivity and passivity indices of nonlinear systems under
  operational limitations using approximations.
\newblock \emph{International Journal of Control}, (just-accepted), 1--20.

\end{thebibliography}


\begin{thebibliography}{16}
\providecommand{\natexlab}[1]{#1}
\providecommand{\url}[1]{\texttt{#1}}
\providecommand{\urlprefix}{URL }
\expandafter\ifx\csname urlstyle\endcsname\relax
  \providecommand{\doi}[1]{doi:\discretionary{}{}{}#1}\else
  \providecommand{\doi}{doi:\discretionary{}{}{}\begingroup
  \urlstyle{rm}\Url}\fi

\bibitem[{Abe et~al.(2016)Abe, Inoue, and Adachi}]{abe2016subspace}
Abe, Y., Inoue, M., and Adachi, S. (2016).
\newblock Subspace identification method incorporated with a priori information
  characterized in frequency domain.
\newblock In \emph{2016 European Control Conference (ECC)}, 1377--1382. IEEE.

\bibitem[{Brogliato et~al.(2007)Brogliato, Lozano, Maschke, and
  Egeland}]{brogliato2007dissipative}
Brogliato, B., Lozano, R., Maschke, B., and Egeland, O. (2007).
\newblock Dissipative systems analysis and control.
\newblock \emph{Theory and Applications}, 2.

\bibitem[{Goethals et~al.(2003)Goethals, Van~Gestel, Suykens, Van~Dooren, and
  De~Moor}]{goethals2003identification}
Goethals, I., Van~Gestel, T., Suykens, J., Van~Dooren, P., and De~Moor, B.
  (2003).
\newblock Identification of positive real models in subspace identification by
  using regularization.
\newblock \emph{IEEE Transactions on Automatic Control}, 48(10), 1843--1847.

\bibitem[{Hoagg et~al.(2004)Hoagg, Lacy, Erwin, and Bernstein}]{hoagg2004first}
Hoagg, J.B., Lacy, S.L., Erwin, R.S., and Bernstein, D.S. (2004).
\newblock First-order-hold sampling of positive real systems and subspace
  identification of positive real models.
\newblock In \emph{Proceedings of the 2004 American control conference},
  volume~1, 861--866. IEEE.

\bibitem[{Kevrekidis et~al.(2015)Kevrekidis, Rowley, and
  Williams}]{kevrekidis2015kernel}
Kevrekidis, I., Rowley, C., and Williams, M. (2015).
\newblock A kernel-based method for data-driven koopman spectral analysis.
\newblock \emph{Journal of Computational Dynamics}, 2(2), 247--265.

\bibitem[{Korda and Mezi{\'c}(2018)}]{korda2018linear}
Korda, M. and Mezi{\'c}, I. (2018).
\newblock Linear predictors for nonlinear dynamical systems: Koopman operator
  meets model predictive control.
\newblock \emph{Automatica}, 93, 149--160.

\bibitem[{Lacy and Bernstein(2003)}]{lacy2003subspace}
Lacy, S.L. and Bernstein, D.S. (2003).
\newblock Subspace identification with guaranteed stability using constrained
  optimization.
\newblock \emph{IEEE Transactions on automatic control}, 48(7), 1259--1263.

\bibitem[{Miller and De~Callafon(2013)}]{miller2013subspace}
Miller, D.N. and De~Callafon, R.A. (2013).
\newblock Subspace identification with eigenvalue constraints.
\newblock \emph{Automatica}, 49(8), 2468--2473.

\bibitem[{Narendra and Parthasarathy(1990)}]{narendra1990identification}
Narendra, K.S. and Parthasarathy, K. (1990).
\newblock Identification and control of dynamical systems using neural
  networks.
\newblock \emph{IEEE Transactions on neural networks}, 1(1), 4--27.

\bibitem[{Okada and Sugie(1996)}]{okada1996subspace}
Okada, M. and Sugie, T. (1996).
\newblock Subspace system identification considering both noise attenuation and
  use of prior knowledge.
\newblock In \emph{Proceedings of 35th IEEE Conference on Decision and
  Control}, volume~4, 3662--3667. IEEE.

\bibitem[{Patra et~al.(1999)Patra, Pal, Chatterji, and
  Panda}]{patra1999identification}
Patra, J.C., Pal, R.N., Chatterji, B., and Panda, G. (1999).
\newblock Identification of nonlinear dynamic systems using functional link
  artificial neural networks.
\newblock \emph{IEEE transactions on systems, man, and cybernetics, part b
  (cybernetics)}, 29(2), 254--262.

\bibitem[{Psichogios and Ungar(1992)}]{psichogios1992hybrid}
Psichogios, D.C. and Ungar, L.H. (1992).
\newblock A hybrid neural network-first principles approach to process
  modeling.
\newblock \emph{AIChE Journal}, 38(10), 1499--1511.

\bibitem[{Sebe(2018)}]{sebe2018sequential}
Sebe, N. (2018).
\newblock Sequential convex overbounding approximation method for bilinear
  matrix inequality problems.
\newblock \emph{IFAC-PapersOnLine}, 51(25), 102--109.

\bibitem[{Williams et~al.(2015)Williams, Kevrekidis, and
  Rowley}]{williams2015data}
Williams, M.O., Kevrekidis, I.G., and Rowley, C.W. (2015).
\newblock A data--driven approximation of the koopman operator: Extending
  dynamic mode decomposition.
\newblock \emph{Journal of Nonlinear Science}, 25(6), 1307--1346.

\bibitem[{Xu et~al.(2002)Xu, Yagoub, Ding, and Zhang}]{xu2002neural}
Xu, J., Yagoub, M.C., Ding, R., and Zhang, Q.J. (2002).
\newblock Neural-based dynamic modeling of nonlinear microwave circuits.
\newblock \emph{IEEE Transactions on Microwave Theory and Techniques}, 50(12),
  2769--2780.

\bibitem[{Zakeri and Antsaklis(2019)}]{zakeri2019passivity}
Zakeri, H. and Antsaklis, P.J. (2019).
\newblock Passivity and passivity indices of nonlinear systems under
  operational limitations using approximations.
\newblock \emph{International Journal of Control}, (just-accepted), 1--20.

\end{thebibliography}

                                                   







\end{document}